\documentclass{article}
\begin{document}\title{Feynman-Kac path integral expansion around   the
upside-down oscillator}
\author{ Z. Haba\\ Institute of Theoretical Physics, University of Wroclaw, \\email:zbigniew.haba @uwr.edu.pl} \maketitle
\begin{abstract}We discuss path integrals for quantum mechanics
with a potential which is a perturbation of the upside-down
oscillator. We express the path integral (in the real time) by the
Wiener measure. We obtain the Feynman integral for perturbations
which are the Fourier-Laplace transforms of a complex measure and
for polynomials of the form $x^{4n}$ and $x^{4n+2} $ ( where $n$
is a natural number). We extend the method to quantum field theory
(QFT) with complex scaled spatial coordinates ${\bf x}\rightarrow
i{\bf x}$. We show that such a complex
 extension of the path integral (in the real time) allows
 a rigorous path integral treatment of a large class of potentials
 including the ones unbounded from below.

\end{abstract}
\section{Introduction} The standard perturbation theory is based
on an expansion of the Hamiltonian around the minimum of the
potential $V$. For an expansion around an extremum we have (choose
$x=0$ as the extremum of $V$ and  $V(0)=0$) \begin{equation}
H=\frac{p^{2}}{2m}+V(x)=
\frac{p^{2}}{2m}+\frac{1}{2}V^{''}(0)x^{2}+...=H_{0} +H_{1} .
\end{equation}
If we have the minimum ( $V^{''}(0)=\omega^{2}>0$) then $H_{0}$ is
the oscillator Hamiltonian. We have the Fock ground state for
$H_{0}$ and the Gell-Mann-Low formula for a perturbative expansion
of correlation functions in the true ground state  in terms of the
correlation functions  in the Fock ground state of $H_{0}$. If
$V^{''}(0)=-\nu^{2}<0$ then $H_{0}$ has no ground state. There is
no distinguished state for a calculation of correlation functions.
We investigate solutions of the Schr\"odinger equation via the
Feynman-Kac formula (in the real time). We show that for a large
class of analytic potentials solutions of the Schr\"odinger
equation can be expressed by the Wiener integral. As an
application of this representation we discuss briefly an estimate
on
 the sojourn time \cite{barton}\cite{guth}: let a
particle be in a state $\psi_{0}$ with its support around $x=0$,
what is the mean time of the particle  stay in the interval
$[-a,a]$?. A related problem in quantum field theory (QFT)
concerns the decay of the false vacuum
\cite{coleman1}\cite{coleman2}. We may expect that the question
can be answered  in the model with the $H_{0}$ evolution treating
$H_{1}$ as a perturbation.

An extension of the method to QFT requires a change of sign of the
gradient term $(\nabla\phi)^{2}\rightarrow -(\nabla\phi)^{2}$(this
can be interpreted as the analytic continuation ${\bf
x}\rightarrow i{\bf x}$ in spatial coordinates). In such a case we
work with the real time but with an Euclidean metric. After the
analytic continuation of the Feynman-Kac path integral some field
theoretic models  can be defined in a rigorous way either
non-perturbatively or via a convergent perturbation theory. We
discuss exponential, trigonometric and polynomial interactions (of
the form $\phi^{4n}$ or $\phi^{4n+2}$, where $n$ is a natural
number ).
 We briefly discuss  the problem of representing the
  time evolution in a Hilbert space. The problem concerns
  an integration  of the wave functions resulting  from the
  Feynman-Kac formula.
 The
analytic continuation to Minkowski space-time in more than two
dimensions encounters difficulties with the convergence of
integrals. Nevertheless, with this approach we have another method
to start with a well-defined field theoretic model which cannot be
approached directly by means of standard methods. The return to
the Minkowski signature may be unnecessary. The change of the
signature $(\nabla\phi)^{2}\rightarrow -(\nabla\phi)^{2}$ may be a
physical process in Einstein gravity
  at
 the Planck scale distances \cite{visser3} \cite{visser1}\cite{visser2}.
 The
use of complex coordinates and    complex scaling has  been
developed in a description of resonances in quantum mechanics (see
the reviews in \cite{cs}\cite{simoncs}).
 The scaling $x\rightarrow
\sqrt{i} x$ is considered in \cite{habajmpqed}\cite{hababook} and
in \cite{am1}\cite{am2}\cite{mazz}. The complex scaling of scalar
fields appeared in some recent papers concerning the functional
integration in lattice QFT \cite{bender}\cite{romatschke}.

The plan of the paper is the following. First, we discuss
perturbations of the upside-down oscillator  in one dimension in
Heisenberg and Schr\"odinger formulation (secs.2-3). Then, we
apply a similarity transformation by means of a solution of the
oscillator equation (sec.4) arriving at a diffusion-like equation
(with an imaginary diffusion constant). In secs.5-7 we show that
for a class of analytic potentials the solution of the transformed
Schr\"odinger equation can be expressed by an expectation value
over a complexified Ornstein-Uhlenbeck process. In sec.8 we
explain that  the method of the upside-down oscillator works in
QFT of a scalar field after an analytic continuation of the
gradient term in the Lagrangian $(\nabla\phi)^{2}\rightarrow
-(\nabla\phi)^{2}$. We demonstrate (sec.9) the effect of this
analytic continuation in the formal loop expansion (in the  real
time) in QFT. Subsequently, we extend the method developed  in
quantum mechanics in secs.2-7 to the path integral in QFT
beginning with fields whose spatial coordinates are put on the
lattice (sec.10). We proceed to the continuum formulation showing
that in the field theory with exponential and trigonometric
interactions (sec.11) the Feynman formula makes sense in
perturbation expansion in any dimension. For polynomial
interactions (sec.12) $ \phi^{4n-2}$ ($n$ is a natural number) the
Feynman-Kac formula in QFT models can be established in two
dimensions. In $\phi^{4n}$ models the Feynman-Kac solution of the
Schr\"odinger equation with the initial condition
$\psi(\sqrt{i}\phi)$ and the inverted signature exists in any
dimension. The return to the Minkowski space-time ${\bf
x}\rightarrow i{\bf x}$  is
                   possible   in two
space-time dimensions.

\section{Upside-down oscillator}
In this section we  consider a quadratic potential ( with
$\omega^{2}=-\nu^{2}<0$) which is unbounded from below.The
Schr\"odinger equation for this oscillator is (the mass $m=1$, $x\in R$)
\begin{equation}
i\hbar\partial_{t}\psi_{t}=(-\frac{\hbar^{2}}{2}\nabla_{x}^{2}-\frac{\nu^{2}
x^{2}}{2})\psi_{t}\equiv H_{0}\psi_{t}.
\end{equation}

 In the Heisenberg picture
\begin{equation} x_{t}=U_{t}^{+}xU_{t},
\end{equation}where $U_{t}$ is the Hamiltonian unitary time
evolution. The model (3) is soluble in the Heisenberg picture
(where $p$ is the momentum operator)
\begin{equation}
x_{t}=x\cosh(\nu t)+\frac{\sinh(\nu t)}{\nu}p.
\end{equation}
We are interested in  the correlation functions
\begin{equation}
(\psi_{0},F_{1}(x_{t})F_{2}(x)\psi_{0}) =
(\psi_{t},F_{1}(x)U_{t}F_{2}(x)\psi_{0})
\end{equation}
and (for the sojourn time)in the probability to stay in an
interval $\Omega$ around the maximum
\begin{equation}
\int_{\Omega}dx\vert \psi_{t}(x)\vert^{2}=(\psi_{t},\theta
(\vert\Omega\vert -\vert x\vert)\psi_{t}),
\end{equation}
where $\vert\Omega\vert$ denotes the length of the interval. We
can approximate the Heaviside step function as $\theta
(\vert\Omega\vert -\vert x\vert)\simeq
\exp(-(\frac{2x}{\vert\Omega\vert})^{2n} )$ with a certain natural
$n$.

We consider a potential
\begin{equation} V=-\frac{\nu^{2}}{2}x^{2}+\tilde{V}(x)
\end{equation}
Then,  the Heisenberg equations of motion are
\begin{displaymath}
(-\frac{d^{2}}{dt^{2}}+\nu^{2})x_{t}=\tilde{V}^{\prime}(x_{t}).
\end{displaymath}
Using the Green function for the operator on the lhs
\begin{displaymath}
G(t,s)=\frac{1}{2\nu}\exp(-\nu\vert t-s\vert)
\end{displaymath}
we rewrite  the Heisenberg equation of motion as an integral
equation
\begin{equation}\begin{array}{l}
x_{t}=x\cosh(\nu t)+\frac{\sinh\nu t}{\nu}p
+\int_{0}^{t}G(t,s)V^{\prime}(x_{s})ds .
\end{array}\end{equation}
Eq.(8) can be solved by iteration. It supplies a  method to
calculate perturbatively  correlation functions (5)-(6) which can
be compared with  calculations by means of other methods.

\section{Dyson perturbation expansion}
 We
consider  the Schr\"odinger equation
\begin{equation}
\partial_{t}\psi=\hat{H}\psi,
\end{equation}
where
\begin{equation}\hat{H}=i\frac{\hbar}{2}\frac{d^{2}}{dx^{2}}
+\frac{i}{\hbar}\nu^{2}x^{2}-\frac{i}{\hbar}\tilde{V}(x) \equiv
\hat{H}_{0}-\frac{i}{\hbar}\tilde{V}(x).\end{equation}

  We can solve the Schr\"odinger equation
(9) perturbatively by means of the Dyson expansion
\begin{equation}
\begin{array}{l}
\psi_{t}=\exp(\tilde{H}_{0}t)\psi
-\frac{i}{\hbar}\int_{0}^{t}ds\exp((t-s)\tilde{H}_{0})\tilde{V}\exp(s\tilde{H}_{0})\psi
\cr -\hbar^{-2}\int_{0}^{t}ds\int_{0}^{s}ds^{\prime}\exp((t-s)
\tilde{H}_{0})\tilde{V}\exp((s-s^{\prime})\tilde{H}_{0})\tilde{V}\exp(s^{\prime}\tilde{H}_{0})\psi
+....
\end{array}
\end{equation}
By direct differentiation we can check that if the series (11) is
uniformly convergent, so that we can exchange differentiation with
the (infinite) sum, then the sum of the series is the solution of
the Schr\"odinger equation. In eq.(11) the propagator follows from
the Mehler formula (\cite{merzbacher} with $\omega\rightarrow
i\nu$)
\begin{equation}\begin{array}{l}
K(t;x,y)=\exp(t\hat{H}_{0})(x,y)\cr=(2\pi\hbar i\nu^{-1}\sinh(\nu
t))^{-\frac{1}{2}} \exp\Big(\frac{i\nu}{2\hbar\sinh (\nu
t)}\Big((x^{2}+y^{2})\cosh(\nu t)-2xy\Big)\Big).\end{array}
\end{equation}
We could calculate the expectation values (5)-(6) for arbitrary
$\psi$  either using the  solution (8) in the Heisenberg picture
or
 calculating the integrals (11) with the evolution kernel (12).

\section{Expansion around a Gaussian  solution}
We discuss  in this section the Schr\"odinger equation (9)

\begin{equation}
i\hbar\partial_{t}\psi_{t}=(-\frac{\hbar^{2}}{2}\nabla_{x}^{2}-\frac{\nu^{2}
x^{2}}{2}+\tilde{V}(x))\psi_{t}\equiv H\psi_{t}
\end{equation}    still
in another approach. We introduce a  solution $\psi_{t}^{\Gamma}$ of the Schr\"odinger
equation for the upside-down oscillator
\begin{equation}
i\hbar\partial_{t}\psi_{t}^{\Gamma}=(-\frac{\hbar^{2}}{2}\nabla_{x}^{2}-\frac{\nu^{2}
x^{2}}{2})\psi_{t}^{\Gamma}.
\end{equation}
We represent the solution of eq.(13) in the form
\begin{equation}
\psi_{t}=\psi_{t}^{\Gamma}\chi_{t}.
\end{equation}
Then, $\chi_{t}$ solves the equation
\begin{equation}
\partial_{t}\chi_{t}=(\frac{i\hbar}{2}\nabla_{x}^{2}+i\hbar
\nabla_{x}\ln\psi_{t}^{\Gamma}\nabla-\frac{i}{\hbar}\tilde{V})\chi_{t}\equiv
\frac{1}{i\hbar}\tilde{H}\chi_{t}
\end{equation}
with the initial condition
$\chi_{0}=\psi_{0}(\psi_{0}^{\Gamma})^{-1}$ determined by the
initial conditions of $ \psi_{t}$ and
 $\psi_{t}^{\Gamma}$.

 Let us consider a particular solution of eq.(14)
\begin{equation}
\psi_{t}^{\Gamma}=\exp(-\frac{\nu}{2}t)\exp( i\frac{\nu
x^{2}}{2\hbar}) .
\end{equation}
Then, eq.(15) reads
\begin{equation} \psi_{t}(x)=\exp(-\frac{\nu}{2}t) \exp(i\frac{\nu
x^{2}}{2\hbar})\chi_{t}(x).
\end{equation}
$\tilde{H}$ in eq.(16) comes from the similarity transformation
\begin{equation}
\tilde{H}=\exp(-i\frac{\nu}{2\hbar}x^{2})H\exp(i\frac{\nu}{2\hbar}x^{2})+\frac{i\nu\hbar}{2}=
-\frac{1}{2}\hbar^{2}\partial_{x}^{2}-i\hbar\nu x\partial_{x}
+\tilde{V} .
\end{equation}

\section{Feynman integral expressed by the Brownian motion}
 With $\tilde{V}=0$ the equation (16) for $\chi$ is a diffusion equation
  with an imaginary diffusion constant and a complex drift. Its
  solution can be expressed by a solution of the Langevin equation
\cite{freidlin}\cite{ikeda}\cite{simonbook}

\begin{equation}
dq_{s}=i\hbar\nabla\ln\psi_{t-s}^{\Gamma}(q_{s})ds+\sigma dw_{s},
\end{equation}where\begin{equation}
\sigma=\sqrt{i\hbar}\equiv
\frac{1}{\sqrt{2}}(1+i)\sqrt{\hbar}.\end{equation} The Brownian
motion $w_{t}$ is the Gaussian process with mean zero and the
covariance
\begin{equation}
E[w_{t}w_{s}]=min(t,s), \end{equation} where $w_{0}=0$ and
$t,s\geq 0$. For the solution (17)  the stochastic equation (20)
reads
\begin{equation}
dq_{s}==-\nu q ds+\sigma dw_{s}.
\end{equation}
 Eq.(23) has the
solution (where $x$ is the initial condition at $t_{0}$)
\begin{equation}
q_{s}(x)=\exp(-\nu(
s-t_{0}))x+\sigma\int_{t_{0}}^{s}\exp(-\nu(s-t))dw_{t}.
\end{equation}
If $t_{0}<0$ then we define $w(t)\equiv \tilde{w}(-t)$ where
$\tilde{w}$ is the Brownian motion independent of the one for
$t\geq 0$.

 We assume that the initial wave function $\chi$ and the
potential $\tilde{V}$ are holomorphic functions. Then, the
solution of eq.(16) is given by the Feynman-Kac formula (assuming
that the expectation value over the Brownian motion on the rhs is
finite)
\begin{equation}
\chi_{t}(x)=E\Big[\exp\Big(-\frac{i}{\hbar}\int_{t_{0}}^{t}ds
\tilde{V}(q_{s}(x))\Big)\chi_{0}(q_{t}( x))\Big],\end{equation}
here $q_{s}(x)$ is the solution (24) of the Langevin equation (23)
with the initial condition $q_{t_{0}}(x)=x$. The solution (25) has
been discussed earlier in \cite{cameron}\cite{doss1}\cite{doss2}
\cite{habajp}\cite{hababook}\cite{habaepj}. It is a real time
version of the Feynman-Kac formula
\cite{freidlin}\cite{simonbook}.

 The formula (25) can be proved  by differentiation (using
the Ito formula \cite{ikeda}\cite{simonbook}). For this purpose
the composition law for the time evolution can be used
$U(t)=U(t-s)U(s)$ which is equivalent to the Markov property (see
\cite{freidlin}). In such a case
 in order to show that the formula (25) solves  eq.(16) it is sufficient to
calculate the generator at $t=0$. From  eq.(25) we have (when
$t_{0}=0$ and $t\rightarrow 0$ )
\begin{displaymath}\begin{array}{l} d\chi_{t}=
E[-\frac{i}{\hbar}V(x)\chi_{0}+\nabla\chi_{0}dq+\frac{1}{2}\nabla^{2}\chi_{0}dqdq].
\end{array}\end{displaymath}
 We insert $dq$ from eq.(23), use $E[Fdw_{s}]=0 $ and $dqdq=i\hbar dt$ (on the
basis of the Ito calculus \cite{simonbook}). After the calculation
of the differentials we may let $t\rightarrow 0$. Then, the rhs of
$d\chi$ is $(\frac{i\hbar}{2}\nabla^{2}-\frac{i}{\hbar}V_{0}-\nu
x\nabla)\chi dt$ (in agreement with  the rhs of eq.(16) at $t=0$).

The solution of the stochastic equation (20) determines the
correlation functions  of the position operator $x_{t}$ in the
Heisenberg picture ( $\tilde{V}=0$)
\begin{equation}\begin{array}{l}(\psi_{0}^{\Gamma},F_{1}(x_{t})F_{2}(x)\psi_{0}^{\Gamma})
=\int dx
\vert\psi_{t}^{\Gamma}(x)\vert^{2}F_{1}(x)E\Big[F_{2}\Big(q_{t}(x)\Big)\Big].
\end{array}\end{equation}
If $\tilde{V}\neq 0$
\begin{equation}\begin{array}{l}(\chi\psi_{0}^{\Gamma},F_{1}(x_{t})F_{2}(x)\chi\psi_{0}^{\Gamma})
\cr=\int dx \vert\psi_{t}^{\Gamma}(x)\vert^{2}F_{1}(x)
E\Big[\exp\Big(-\frac{i}{\hbar}\int_{0}^{t}ds\tilde{V}(q_{s}(x))\Big)\chi\Big(q_{t}(x)\Big)\Big]^{*}\cr\times
E\Big[\exp\Big(-\frac{i}{\hbar}\int_{0}^{t}ds\tilde{V}(q_{s}(x))\Big)F_{2}\Big(q_{t}(x)\Big)\chi\Big(q_{t}(x)\Big)\Big].
\end{array}\end{equation}We write eqs.(26)-(27) for a general
solution $\psi_{t}^{\Gamma}$ of eq.(14). In the particular case
(17) $\vert\psi_{t}^{\Gamma}\vert=\exp(-\frac{\nu}{2}t)$ is $x$-independent. If $\psi_{t}^{\Gamma}$ is in
$L^{2}(dx)$ then we may consider polynomial $F$'s in
eqs.(26)-(27). If $\psi_{t}^{\Gamma}$ is given by eq.(17) then
$F_{1}$ or $F_{2}$ must decay for a large $x$ if the expectation
values (26) are to be finite.

  By
explicit calculation we establish  (it follows also from eq.(25))
the following formula ($t_{0}=0$)
\begin{equation}
\exp(-\frac{\nu t}{2}+\frac{i\nu x^{2}}{2})E[\chi(q_{t}(x))] =\int
K(t;x,y)\exp(\frac{i\nu y^{2}}{2})\chi(y) dy
=(U_{t}\psi)(x)\end{equation} where $K(t;x,y)$ is the Mehler
kernel for the oscillator with $\omega =i\nu$  (see eq.(12)). In
this sense the lhs of eq.(28) gives a path integral solution of
the upside-down oscillator. We do not need to explore whether the
oscillatory Feynman path integral is mathematically well-defined
because the solution (28) is expressed by the Wiener integral. For
the proof of eq.(28) we represent $\chi(x)$ by Fourier transform.
Then, it is sufficient to calculate  both sides of eq.(28) for
$\chi(x)=\exp(ipx)$. On the lhs of eq.(28) we need to compute
\begin{equation}
E\Big[\exp\Big(i\sigma
p\int_{0}^{t}\exp(-\nu(t-s))dw_{s}\Big)\Big]= \exp\Big(-\frac{i
\hbar p^{2}}{4\nu}(1-\exp(-2\nu t))\Big)
\end{equation}

 Using $E[(\int fdw_{s})^{2}]=\int f^{2}ds$ we calculate from
 eq.(24)
\begin{equation}\begin{array}{l}
E[q_{t}(x)q_{t^{\prime}}(x)]\cr=\exp(-\nu (t+
t^{\prime}-2t_{0}))x^{2}+ \frac{i\hbar}{2\nu}\Big(\exp(-\nu\vert
t-t^{\prime}\vert)-\exp(-\nu(t+t^{\prime}-2t_{0}))\Big).\end{array}\end{equation}
We have
\begin{equation}
\lim_{t_{0}\rightarrow -\infty}E[q_{t}(x)q_{t^{\prime}}(x)]
=\frac{i\hbar}{2\nu}\exp(-\nu\vert t-t^{\prime}\vert).
\end{equation}

When $t_{0}=-\infty$ then
\begin{displaymath}
q_{s}=\sigma\int_{-\infty}^{s}\exp(-\nu (s-t))dw_{t}
\end{displaymath}
is independent of $x$ and at $t_{0}\rightarrow -\infty$ (in
eq.(30))
\begin{displaymath}
E[q_{t}(x)q_{t^{\prime}}(x)] = \frac{i\hbar}{2\nu}\exp(-\nu\vert
t-t^{\prime}\vert).
\end{displaymath}

The lhs of eq.(28) is defined for $t\geq 0$ whereas the rhs is the
upside-down oscillator evolution for any time. If $t<0$ then the
corresponding formula for the lhs can be obtained from the
time-reflection relation
\begin{equation}
\psi_{t}=\Big(\exp(iHt)\psi^{*}\Big)^{*}
\end{equation}
which gives the formula
\begin{equation}
\exp(-\frac{\nu t}{2}+\frac{i\nu
x^{2}}{2})E[\chi(\tilde{q}_{t}(x))] =\int K(t;x,y)\exp(\frac{i\nu
y^{2}}{2})\chi(y) dy ,
\end{equation}
where for $t=-\tilde{t}\leq 0$ we define
$\tilde{q}_{t}=\hat{q}(\tilde{t})$ which satisfies the equation
\begin{equation}
d\hat{q}(\tilde{t})=-\nu \hat{q}(\tilde{t})d\tilde{t}+\sigma
dw(\tilde{t})
\end{equation}
with the solution
\begin{equation}
\hat{q}_{\tilde{t}}(x)=\exp(-\nu
\tilde{t})x+\sigma\int_{0}^{\tilde{t}}\exp(-\nu(\tilde{t}-s))dw_{s}.
\end{equation}
Hence, for  $t<0$
\begin{equation}
\tilde{q}_{t}(x)=\exp(\nu
t)x+\sigma\int_{0}^{-t}\exp(\nu(t+s))dw_{s}.
\end{equation}

For a restricted class of potentials the formula (25) or the one
for an analytically continued wave function $\chi(x)\rightarrow
\chi(\sigma x)$ can have a non-perturbative meaning
\begin{equation}
\chi_{t}(\sigma
x)=E\Big[\exp\Big(-\frac{i}{\hbar}\int_{0}^{t}ds\tilde{V}(q_{s}(\sigma
x))\Big)\chi_{0}(q_{t}( \sigma x))\Big].\end{equation}When we
change coordinates in eq.(24) $x\rightarrow \sigma x $

\begin{equation}
q_{s}(\sigma x)=\sigma \xi_{s}(x),
\end{equation} where  $\xi$ is real. For $\nu=0$ we have
$\xi_{s}(x)=x+w_{s}$ whereas for the upside-down oscillator
\begin{equation}
\xi_{t}(x)=\exp(-\nu t)x+\int_{0}^{t}\exp(-\nu(t-s))dw_{s} .
\end{equation}
$\xi_{s}(x)$ is the Ornstein-Uhlenbeck process starting from $x$
at $t=0$ \cite{chandra}.
\section{Potentials which are Fourier-Laplace transforms of a
measure}

We consider potentials of the form of the Fourier-Laplace
transforms of a complex measure

\begin{equation}
\tilde{V}(x)=g\int d\mu(a)\exp(i\alpha ax),
\end{equation}
and wave functions of the same form
\begin{equation}
\psi(x)=\int d\rho(a_{0})\exp(i\alpha a_{0}x),
\end{equation}where $a\in R$.
We consider either $\alpha=1$ or $\alpha=i$. We need
$\mu(a)=\mu(-a)^{*} $ for  $\alpha=1$ and $\mu(a)=\mu(a)^{*} $ for
$\alpha=i$ if the potential is to be real. If $\mu(a)=\mu_{1}(a)
+i\mu_{2}(a)$, where $\mu_{j}$ are real, then we define the norm
$\vert \mu\vert =\vert\mu_{1}\vert+ \vert\mu_{2}\vert$, where
$\vert \mu_{j}\vert $ is the variation of the measure on $R$ so
that $\vert\int fd\mu\vert \leq \vert\mu\vert \sup\vert f\vert $.
The potentials which are Fourier transforms of a measure (with
$\alpha=1$) have been studied in the approach to the Feynman
integral in terms of the Fresnel integral in ref.\cite{ahfeyn}.

In the set of potentials (40) we consider the ones which lead to a
convergent perturbative expansion and we distinguish potentials
with a bounded Feynman-Kac factor.
 As an example of the latter class we consider potentials of the form
\begin{equation}
\tilde{V}(x)=\int_{-\infty}^{\infty} da v(a)\exp(i\alpha ax),
\end{equation}
where
\begin{equation}
v(a)=\int_{0}^{\infty}du f(u)\exp(-\frac{1}{2}a^{2}u),
\end{equation} where $f$ is a real function (or distribution).
Then,
\begin{equation}\begin{array}{l}
\tilde{V}( x)=\int_{0}^{\infty}du f(u)(\frac{u}{2\pi}
)^{-\frac{1}{2}}\exp(-\frac{\alpha^{2}}{2u}x^{2})\end{array}
\end{equation}and
\begin{equation}\begin{array}{l}
\tilde{V}(\sigma x)=\int_{0}^{\infty}du f(u)(\frac{u}{2\pi}
)^{-\frac{1}{2}}\exp(-\frac{i\alpha^{2}\hbar}{2u}x^{2}).\end{array}
\end{equation}
If
\begin{equation}\begin{array}{l}
\Big\vert\exp\Big(-\frac{i}{\hbar}\int_{0}^{t}ds V(q_{s}(\sigma
x))\Big)\Big\vert^{2}\cr =\exp\Big(
-\frac{2}{\hbar}\int_{0}^{t}ds\int_{0}^{\infty} du
f(u)(\frac{u}{2\pi}
)^{-\frac{1}{2}}\sin(\frac{\hbar\alpha^{2}}{2u}\xi_{s}^{2})\Big)\leq
R
\end{array}\end{equation} with a certain function $R>0$ bounded in
$\xi$, then the Feynman-Kac integral (37) is well-defined for
$E[\vert\chi(\sigma\xi_{t})\vert]<\infty$ . If additionally
\begin{equation}
\vert V(x)\vert\leq \int_{0}^{\infty}du \vert f(u)\vert
u^{-\frac{1}{2}}<\infty
\end{equation}
then $\tilde{V}( x) $ for $\alpha=1$ as well as $\tilde{V}(\sigma
x) $ for any $\alpha$ are bounded functions of $x$.

As an example we could consider $f(u)=\frac{1}{2}
\exp(-\frac{1}{2u})u^{-\frac{3}{2}}$. Then,  from eq.(44) with
$\alpha=1$ we obtain the meromorphic potential
$\tilde{V}(x)=(1+x^{2})^{-1}$
 ( the Feynman-Kac formula for meromorphic
potentials is discussed in \cite{hababook}). With $q_{s}(\sigma
x)=\sigma\xi_{s}(x)$  in the Feynman formula (37) the estimate of
the lhs of eq.(46) is (we could obtain also a bound on the Feynman
formula (25) but the argument is more involved)
\begin{displaymath}
\exp\Big(-\int_{0}^{t}ds\frac{i}{1+i\hbar\xi^{2}}+\int_{0}^{t}ds\frac{i}{1-i\hbar\xi^{2}}\Big)
=\exp\Big(-\int_{0}^{t}ds\frac{2\xi^{2}}{1+\xi^{4}}\Big) .
\end{displaymath}
It follows
\begin{displaymath}
\Big\vert \exp\Big(-\frac{i}{\hbar}\int_{0}^{t}V(q_{s}(\sigma
x))ds\Big)\Big\vert^{2}\leq R(t)
\end{displaymath}
where the rhs does not depend on $x$ and $\xi_{s}$. Then, the
Feynman integral (37) is defined for all $\chi$ (41) with
$\vert\rho\vert<\infty$ .

 We are going to prove for a larger class of
functions that the solution of eq.(13) can be expressed  by the
formula (25) as a convergent perturbation series

\begin{equation}
\chi_{t}(x)=E\Big[\sum_{n}\frac{1}{n!}\exp\Big(-\frac{i}{\hbar}\int_{t_{0}}^{t}\tilde{V}(q_{s})ds\Big)^{n}\chi_{0}(q_{t}(x))\Big],
\end{equation} where $\tilde{V}$ has the representation (40) and $\chi$ the
representation (41) \newline $\chi_{0} =\int
d\rho(a_{0})\exp(i\alpha a_{0} x)$. We do not prove that
$\Big\vert\exp\Big(-\frac{i}{\hbar}\int_{t_{0}}^{t}\tilde{V}(q_{s})ds\Big)\Big\vert$
has a finite expectation value ( except of the special class of
functions (44)-(45)). For this purpose we would need some cutoffs
(as in \cite{habajmp}\cite{abh}) which subsequently are removed in
the perturbation series. We show that the perturbation series of
the formula (48) in powers of $\tilde{V}$ is absolutely convergent
(the perturbation series (48) and the Dyson series (11) coincide,
see \cite{habauni}\cite{habaspringer}). Then, the series can be
differentiated term by term. As a consequence the sum of the
series gives the solution of the Schr\"odinger equation (13). The
details of the proof are the same as in \cite{habajmp}\cite{abh}
(where $\nu=0$ is considered). In order to prove the convergence
of the perturbation series we calculate the expectation value of
the $N$-th order term. We perform the calculation for $\nu>0$ in
eq.(24) . The special case
 $\nu=0$ is discussed in \cite{habajmp} (it also follows as a limit of  the formulas below).
  When we apply the Fourier-Laplace representation of $\chi$ and
$\tilde{V}$ then we can see that the $N$-th order term is of the
form (we skip the time integration in front of the $N$-th order
term and set $t_{0}=0$ )
\begin{equation}\begin{array}{l}
\int d\rho (a_{0})
\Pi_{j}d\mu(a_{j})\exp(\sum_{\gamma,\beta}f(a_{\gamma},a_{\beta}))\exp
\Big(-\frac{1}{2}\sigma^{2}\alpha^{2}\cr\times\sum_{\gamma,\beta}a_{\gamma}a_{\beta}
E\Big[\int_{0}^{s_{\beta}}
\exp(-\nu(s_{\beta}-s))dw_{s}\int_{0}^{s_{\gamma}}
\exp(-\nu(s_{\gamma}-s^{\prime}))dw_{s^{\prime}}\Big]\Big)\cr
=\int d\rho(a_{0})
\Pi_{j}d\mu(a_{j})\exp(\sum_{\gamma,\beta}f(a_{\gamma},a_{\beta}))\cr
\times\exp
\Big(-\frac{1}{2}\sigma^{2}\alpha^{2}\sum_{\gamma,\beta}a_{\gamma}a_{\beta}
\int_{0}^{min(s_{\gamma},s_{\beta})}\exp(-\nu(s_{\beta}+s_{\gamma}-2s))ds\Big)
\cr\equiv \int d\mu d\rho\exp(f)\exp(F),\end{array}
\end{equation}where the indices $\gamma,\beta$ are either  $0$ or $j$. In eq.(49) we used the
formula
\begin{displaymath}
E[\exp g(w)]=\exp(\frac{1}{2}E[g(w)^{2}])
\end{displaymath}
true if $E[g]=0$ and $g$ is a linear functional of $w$.
  We have
\begin{equation}
f(a_{j},a_{k})=i\alpha a_{j} x\exp(-\nu s_{j})+i\alpha a_{k}
x\exp(-\nu s_{k})
\end{equation}
and
\begin{equation}
f(a_{0},a_{k})=i\alpha a_{0}x\exp(-\nu t) +i\alpha a_{k}
x\exp(-\nu s_{k}).
\end{equation}  The term $\exp F$ in eq.(49) is a pure phase ( as
$\sigma^{2}=i\hbar$)
\begin{displaymath}\begin{array}{l}\Big\vert \exp
\Big(-\frac{1}{2}\sigma^{2}\alpha^{2}\sum_{\gamma\beta}a_{\beta}a_{\gamma}
\int_{0}^{min(s_{\beta},s_{\gamma})}\exp(-\nu(s_{\beta}+s_{\gamma}-2s))ds\Big)\Big\vert^{2}=1
\end{array}
\end{displaymath}
The $N$-th order term is bounded by
\begin{equation}
\int_{0}^{t}ds_{N}...\int_{0}^{s_{2}}ds_{1}d\vert\rho\vert(a_{0})\prod_{j}d\vert\mu\vert(a_{j})\vert\exp(f)\vert
\end{equation}
If $\alpha=1$ (Fourier case) then $f$ is purely imaginary and the bound (52)
is
\begin{displaymath}
\frac{t^{N}}{N!}\vert\rho\vert \vert\mu\vert^{N}
\end{displaymath}
If $\alpha =i$ then denote
\begin{displaymath}
\hat{V}(\exp(-\nu s) x)=\int d\vert\mu\vert(a) \exp(-a\exp(-\nu s)
x)
\end{displaymath}
From eq.(52) we obtain the bound

\begin{displaymath}
\vert \int_{0}^{t}ds_{N}...\int_{0}^{s_{2}}ds_{1}\int
d\vert\rho\vert(a_{0}) \prod_{j}\hat{V}(\exp(-\nu s_{j})
x)\exp(-a_{0}\exp(-\nu t)x).
\end{displaymath}
If $\hat{V}(e^{-\nu s}x)$ is bounded by $B(x)$ then the term (52) is
bounded by $\frac{t^{N}}{N!} B(x)^{N}A_{t}(x)$ where
\begin{displaymath}
A_{t}(x)=\int d\vert\rho\vert(a_{0})\exp(-a_{0}e^{-\nu
t}x)\end{displaymath}  It follows that the series
(48) under above mentioned assumptions is absolutely convergent.
When we differentiate eq.(48) over $t$ then the resulting series
is of the same form as (49). Hence, the series of derivatives is
also absolutely convergent. The same argument applies when
calculating $\tilde{H}\chi_{t}$. It follows that the sum of the
series (48) is the solution of eq.(48) ( the detailed proof is the
same as in \cite{abh} with $\nu=0$ ).

In this way we show that the expansion of the Feynman-Kac integral
(25) in $\tilde{V}$ leads to a perturbative solution because the
series is absolutely convergent, the differentiation of eq.(48)
over $t$ proves that after the differentiation the lhs of eq.(48)
is convergent and is equal to $\frac{1}{i\hbar}\tilde{H}\chi_{t}$
proving eq.(16).

If we let (as in eq.(37)) $x\rightarrow \sigma x$ then the
integrals over the initial values in eq.(52) change (
$x\rightarrow \sigma x$ in $f$ ) leading to  a minor change of the
class of potentials $\tilde{V}$ and initial wave functions $\chi$
which admit a convergent perturbation series.

\section{Polynomial potentials}
In this section we investigate  the
factor \begin{equation}
\exp\Big(-i\int_{0}^{t}\tilde{V}(q_{s}(x))ds\Big)
\end{equation}
in the Feynman-Kac formula  as a function of $\xi$.  Let us denote
$q_{s}$ of eq.(24) ($t_{0}=0$) as
\begin{equation} q_{s}(x)=\exp(-\nu s)x +\sigma b_{s}
\end{equation}
where $b_{s}=\xi_{s}(0) $  is a real random function (the
Ornstein-Uhlenbeck process starting from 0). The function (53) is
explicitly bounded in $b$ for polynomial potentials
\index{polynomial interaction} with the highest order term
$gx^{2n}$ if
\begin{equation}
-ig\sigma^{2n}=-gi^{n+1}<0.
\end{equation}
Hence, if $n+1=2k$ , where $k$ is a natural number
($\tilde{V}=gx^{4k-2}$), then the condition (55) is
$-g(-1)^{k}<0$. So $k$ should be even if $g>0$ and $k$ odd if
$g<0$. For $k=1$ we have $V=-\nu^{2}x^{2} $ .

As a typical example  of a bounded exponential (53) we may consider ($g>0$)
\begin{displaymath}
\tilde{V}(x)=gx^{6}+\lambda x^{4}+\frac{1}{2}\omega^{2}x^{2}.
\end{displaymath}
Then
\begin{displaymath}\begin{array}{l}
\Big\vert\exp\Big(-i\int_{0}^{t}ds\tilde{V}(\exp(-\nu s)x+\sigma
b(s))\Big)\Big\vert\cr = \exp\Big(-\int_{0}^{t}ds
(gb(s)^{6}+p(x,b(s)))\Big) \leq
R(t,x),\end{array}\end{displaymath} where $p(x,b(s)) $ is a real
polynomial of the fifth order and $R(t,x)$ is a function which
does not depend on $b$. The bound holds true because
$gb^{6}+p(x,w) \geq -\gamma(x)$ for a certain function
$\gamma(x)$. In such a case the expectation value (25) is
integrable if
\begin{displaymath}
E[\vert \chi(\exp(-\nu t)x+\sigma b(t))\vert]<\infty.
\end{displaymath}
e.g., for $\chi$ of the form (41) with $\alpha=1$ and
$\vert\rho\vert<\infty$ .  In the Appendix we discuss briefly a
rough estimate on the sojourn time in the $x^{6}$ potential. Then,
the decay of $q_{s}(x)$ in eq.(24) means large probability to find
a particle far from zero (depending on the potential barrier
$x^{6} $) .

An analytic continuation of coordinates of the type (37) leads to
a larger class  of potentials which can be treated by means of the
Feynman integral represented by the Wiener measure. If in eq.(25) $\tilde{V}(\exp(-\nu
s)x+\sigma b(s))$ is replaced by $\tilde{V}\Big(\sigma(\exp(-\nu s)x+
b(s))\Big)\equiv \tilde{V}(\sigma\xi_{s})$ (eq.(37)) then
\begin{equation}
E\Big[\Big\vert\exp\Big(-i\int_{0}^{t}\tilde{V}(\sigma \xi_{s}(
x))\Big)\Big\vert\Big]<\infty
\end{equation}
 for a subclass of polynomials of the form
\begin{equation}
\tilde{V}(x) =a x+\sum_{n=1}^{N} c_{2n}x^{2n},
\end{equation}
where $a$ is an arbitrary real number and $n$ is a natural number.
If $ n=2k$ (where $k$ is a natural number) then $\sigma^{4k}$ is
real and the exponential in eq.(56)  is a pure
phase,$\vert\exp(i\int(\sigma b)^{2n})\vert =1$, so $c_{4k}$ can
be an arbitrary real number. If $n$ is odd and
$ic_{2n}\sigma^{2n}>0$ then $\exp(-ic_{2n}\sigma^{2n}\int
b^{2n})<1$. Hence, in the Feynman-Kac formula (37) the exponential
is bounded for potentials of the form $x^{4k-2}$ ($x^{6}$ for
$k=2$). As an example we could consider the potential $V(x)=
c_{2}x^{2}+c_{4}x^{4}$ where $c_{2}<0$ and $c_{4}\in R$. Then, the
expression (56) is bounded by
\begin{displaymath}
E[\exp(c_{2}\int b^{2})]<1
\end{displaymath}
if $c_{2}<0$.

 If the highest order term  $N=2k+1$ and in the sum (57) we consider powers $x^{2n}$ with $n<N$ then we need for
the boundedness  of the factor (53) or (56)
 $c_{2N}\sigma^{4k+2}i>0$. Hence, $c_{2N}
(-1)^{k}<0$. If the term with the  highest power $c_{2N}x^{2N}$
($\sim x^{4k-2}$ ) of the potential (57) is negative  then the
Hamiltonian is not essentially self-adjoint \cite{reedsimon1}. In
such a case the self-adjoint extension defining the unitary
evolution is not unique.
 The
Feynman path integral can determine the choice of the extension.
Such an interpretation of the Feynman integral has been suggested
by Nelson \cite{nelson} in a discussion of singular potentials
whose Hamiltonians fail to be essentially self-adjoint. If
$\psi_{t}$ is the solution of the Schr\"odinger equation then by
time differentiation we can see that the scalar product is
preserved $(\psi_{t},\psi_{t})=(\psi_{0},\psi_{0})$ if
\begin{displaymath}
\int_{-\infty}^{\infty}dx\frac{d}{dx}(\psi_{t}^{*}\frac{d}{dx}\psi_{t}
-\psi_{t}\frac{d}{dx}\psi_{t}^{*})=0
\end{displaymath}
For the solution (18) this condition demands $ x\nu\vert
\chi_{t}\vert^{2}\rightarrow 0$ when $\vert x\vert \rightarrow
\infty$ and
$(\chi_{t}^{*}\frac{d}{dx}\chi_{t}-\chi_{t}\frac{d}{dx}\chi_{t}^{*})(\infty)-
(\chi_{t}^{*}\frac{d}{dx}\chi_{t}-\chi_{t}\frac{d}{dx}\chi_{t}^{*})(-\infty)=0$
as a requirement for the domain of definition of the self-adjoint
Hamiltonian.

There remains to discuss the complexification of coordinates in the wave functions
$\psi_{t}(\sigma x)$ in quantum mechanics ( we follow the argument of ref.\cite{cs} from the theory of resonances).
 We are usually interested in a calculation of
transition amplitudes
\begin{displaymath}\begin{array}{l}
(\psi^{\prime},U_{t}\psi)=\int_{-\infty}^{\infty}
dx\psi^{\prime*}(x)\psi_{t}(x)=\int_{-\infty}^{0}
dx\psi^{\prime*}(x)\psi_{t}(x)+\int_{0}^{\infty}
dx\psi^{\prime*}(x)\psi_{t}(x).\end{array}
\end{displaymath}
 On the basis of the Cauchy theorem on an integral of analytic functions we can
replace the integral from the real line to the remaining part of a
closed contour. So for the potentials satisfying the condition
(55) (as $x^{6}$) we can write
\begin{equation}\begin{array}{l}
\int_{0}^{\infty}
dx\psi^{\prime*}(x)\psi_{t}(x)=-\sigma\int_{0}^{\infty}
dx\psi^{\prime*}(\sigma x)\psi_{t}(\sigma x) \cr-\lim_{R\rightarrow
\infty}
iR\int_{0}^{\frac{\pi}{4}}d\theta\psi^{\prime*}(Re^{i\theta})\psi_{t}(Re^{i\theta})
\end{array}\end{equation}
and a similar expressions for the integral on $[-\infty,0]$. Both
integrals on the rhs of eq.(58) exist for the potentials (55).
They may be more efficient for calculations with the Feynman
integral than the one on the lhs. It can be difficult to give a
general rule for the choice of integration contours but in a
particular case such a procedure may be fruitful. The extension of
the wave functions to a complex domain discussed in this paper may
be related to some recent investigations on a definition of path
integrals in a complex domain of lattice field theories
\cite{bender}\cite{romatschke}.
\section{Upside-down oscillator in quantum field theory}
In the paper of Guth and  Pi \cite{guth} the upside-down
oscillator in quantum mechanics  has been discussed as a model for
an exponential expansion from the false ground state. It has been
generalized there to the Higgs model ( with the $-\mu^{2}$ mass
term) in the background of the de Sitter metric. The exponential
expansion of fields in this model is preserved (for a large time)
in the form known from quantum mechanics of secs.1-7 because in de
Sitter space the kinetic term $\triangle \phi$ in the wave
equation is multiplied by the exponentially decreasing factor
$\exp(-2{\cal H}t)$ , where ${\cal H}$ is the Hubble constant. We
show in this section that the field's exponential expansion fails
in Minkowski space in the model with the Lagrangian

\begin{displaymath}
 {\cal
 L}=\frac{1}{2}((\partial_{0}\phi)^{2}-(\nabla\phi)^{2}+\mu^{2}\phi^{2}).
 \end{displaymath}
In the Heisenberg picture the equation of evolution of the quantum
field reads ( for Fourier components)
\begin{equation}
\partial_{t}^{2}\tilde{\phi}_{t}({\bf k})+({\bf k}^{2}-\mu^{2})\tilde{\phi}_{t}({\bf
k})=0 .
\end{equation}
It will have the solution of the form (4) (exponentially
expanding)  for the modes with ${\bf k}^{2}<\mu^{2}$. However, the
modes with ${\bf k}^{2}>\mu^{2}$ will stay at the ``false vacuum"
and oscillate around zero.

We look for Gaussian solutions of the Schr\"odinger equation in
the form
\begin{equation}\begin{array}{l}
\psi_{t}^{\Gamma}=A(t)\exp\Big(-\frac{1}{2\hbar}\int d{\bf x}d{\bf
y}\phi({\bf x})\Gamma_{t}({\bf x}-{\bf y})\phi({\bf y})\Big) \cr=
A(t)\exp\Big(-\frac{1}{2\hbar}\int d{\bf k}\tilde{\phi}(-{\bf
k})\tilde{\Gamma}_{t}({\bf k})\tilde{\phi}({\bf k})\Big) .
\end{array}\end{equation}
$ \psi_{t}^{\Gamma}$ satisfies the Schr\"odinger equation if
$\tilde{\Gamma}$ satisfies the equation (see
\cite{habauni}\cite{habaspringer})\begin{equation}
i\partial_{t}\tilde{\Gamma}-\tilde{\Gamma}^{2}+{\bf
k}^{2}-\mu^{2}=0 .
\end{equation}
If we introduce
\begin{equation}
u_{t}=\exp(i\int_{0}^{t}ds\tilde{\Gamma}_{s})
\end{equation}
then $u$ satisfies a linear equation
\begin{equation}
\partial_{t}^{2}u+({\bf
k}^{2}-\mu^{2})u=0  .
\end{equation}
$\tilde{\Gamma}_{t}$ is obtained from $u$ as
\begin{equation}
i\tilde{\Gamma}_{t}=u_{t}^{-1}\partial_{t}u .
\end{equation}
Eq.(63) coincides with the Heisenberg equation (59). Hence, in the
Schr\"odinger picture we have the same problem as in the
Heisenberg picture: different behavior of the ${\bf
k}^{2}>\mu^{2}$ and  ${\bf k}^{2}<\mu^{2}$ modes. If we put the
field $\phi({\bf x})$ in a box with periodic boundary conditions
then the ${\bf k}$ integral in the exponential of eq.(60) becomes
a sum and the solution (60) is a product of one-dimensional
solutions (analogs of the oscillators of secs.1-7 with real or
imaginary $\nu$ depending on wether ${\bf k}^{2}<\mu^{2}$ or
 ${\bf k}^{2}>\mu^{2}$) .

In order to preserve the analogy with the upside-down oscillator
we need the Lagrangian ( with the potential $\tilde{V}$)
 \begin{equation}
 {\cal
 L}=\frac{1}{2}((\partial_{0}\phi)^{2}+(\nabla\phi)^{2}+\mu^{2}\phi^{2})-\tilde{V}(\phi)
 \end{equation}
 which is Euclidean invariant. Then, the energy density
\begin{displaymath}
 H(x)=\frac{1}{2}((\partial_{0}\phi)^{2}-(\nabla\phi)^{2}-\mu^{2}\phi^{2})+\tilde{V}(\phi)
 \end{displaymath} is  Lorentz invariant. Such a model could be
 used for a description in a real time of the theories
 with negative potentials .
We could treat it as a path integral version for calculations
which finally will need an analytic continuation to Minkowski
space-time. In such a case we could generalize  the model (65)
considering the replacement
\begin{displaymath}
(\nabla\phi)^{2}\rightarrow \zeta^{2}(\nabla\phi)^{2}
\end{displaymath} and $\mu^{2}\rightarrow \zeta^{2}\mu^{2}$ where $\zeta=\alpha+i\beta$ with
$\Re\alpha\geq 0$ (the Minkowski version corresponds to
$\alpha\rightarrow 0$ and $\beta \rightarrow 1$).
In such a case eq.(63) takes the form (eq.(64) remains unchanged)
\begin{equation}
\partial_{t}^{2}u-\zeta^{2}({\bf
k}^{2}+\mu^{2})u=0
\end{equation}          Then,
$\nu\rightarrow \zeta\nu $ ( $\nu=\sqrt{-\triangle +\mu^{2}}$) and
in eq.(24) $\exp(-t\zeta\nu)$ is still a bounded semigroup. The
stochastic process  with $\nu\rightarrow i\omega$ ($\zeta=i$) has
been discussed in \cite{habaspringer}\cite{habaepj}.The Lagrangian
(65) could also appear in path integral
  calculations when the complex saddle points are
needed as for example in the problem encountered in the
description of the tunnelling. In   models with quantized gravity
the metric signature inversion is discussed as a physical
phenomenon
\cite{visser1}\cite{visser2}\cite{hawking}\cite{sorkin}. Then, the
Lagrangian (65) would describe the sector (possibly at the Planck
scale  distances) with the reversed signature. In refs.
\cite{visser3}\cite{visser1}\cite{visser2}  a transition between
sectors with different signatures is discussed similar to our
preliminary consideration in this section. If we consider the
dynamical process of signature change (either in classical or
quantum gravity) then we should discuss a time dependent $\zeta$.
The dynamics of the scalar field as a function of the signature
$\zeta$ is an interesting problem in the theory of differential
equations \cite{visser1}. In any case without the signature change
(possibly with a time dependent $\zeta$ ) the exponentially fast
expansion of all the momentum components of the scalar field
$\tilde{\phi}({\bf k})$ is not possible.

\section{Formal $ \hbar$ expansion in the Feynman path integral in QFT}

We  calculate the generating functional of the model (65) in a
formal expansion in $\hbar$ (till the $O(\sqrt{\hbar})$ terms
\cite{coleman} )\begin{equation}
\begin{array}{l}Z[u]=\int d\phi\exp(\frac{i}{\hbar}\int dx({\cal
L}+u\phi))\cr= \exp\Big(\frac{i}{\hbar}\int dx ({\cal
L}(\phi_{c})+u\phi_{c})\Big) \det\Big(i(
-\partial_{t}^{2}-\nabla^{2}+\mu^{2}-V^{''}(\phi_{c}))\Big)^{-\frac{1}{2}},
\end{array}\end{equation}where
$\phi_{c}(t,{\bf x})\equiv \phi^{c}_{t}({\bf x})$ is the solution
of the equation
\begin{equation}
(-\partial_{t}^{2}-\nabla^{2}+\mu^{2})\phi_{c}-\tilde{V}^{\prime}(\phi_{c})=-u .
\end{equation}
For the propagator we have
\begin{equation}\begin{array}{l}
K(t;\phi,\phi^{\prime})=\int_{\phi_{0}=\phi,\phi_{t}=\phi^{\prime}}
d\phi\exp(\frac{i}{\hbar}\int dx{\cal L})\cr=
\exp\Big(\frac{i}{\hbar}\int dx {\cal L}(\phi_{c})\Big)\cr\times
\det\Big(i(
-\partial_{t}^{2}-\nabla^{2}+\mu^{2}-\tilde{V}^{\prime\prime}(\phi_{c}))\Big)^{-\frac{1}{2}},
\end{array}\end{equation}where
\begin{equation}
(-\partial_{t}^{2}-\nabla^{2}+\mu^{2})\phi_{c}-\tilde{V}^{\prime}(\phi_{c})=0.
\end{equation}
Eq.(70) is solved with the boundary conditions
$\phi^{c}_{0}=\phi,\phi^{c}_{t}=\phi^{\prime}$ . The equations
(68) and (70) have a  form similar to the ones in Euclidean field
theory but the potential enters with an opposite sign.

For $\tilde{V}=0$ we can obtain explicit formulae from
eqs.(67)-(70)
\begin{equation}
Z[u]=\exp(-\frac{1}{2\hbar}uGu) ,\end{equation}
 where
\begin{equation}
G(t,{\bf x};t^{\prime},{\bf x}^{\prime})=i\Big(\exp (-\nu\vert
t-t^{\prime}\vert)(2\nu)^{-1}\Big)({\bf x},{\bf x}^{\prime})\equiv
iG^{E}(t,{\bf x};t^{\prime},{\bf x}^{\prime})
\end{equation}
where  $G^{E}$ is the two-point function for the Euclidean free
field (with $\nu=\sqrt{-\triangle +\mu^{2}}$). The evolution
kernel calculated from eq.(69) is
\begin{equation}\begin{array}{l}
K(t;\phi,\phi^{\prime})=\det\Big(2\pi i\nu^{-1}\sinh (\nu
t)\Big)^{-\frac{1}{2}}\exp\Big(\frac{i}{2}\phi \nu\coth(\nu t)\phi
\cr+\frac{i}{2}\phi^{\prime} \nu\coth(\nu
t)\phi^{\prime}-\phi\frac{i\nu}{\sinh(\nu
t)}\phi^{\prime}\Big)\end{array}
\end{equation}
The formula (73) can be made rigorous  if we put the spatial
coordinates on the lattice (this will be discussed in the next
section). In another approximation to eq.(69) we could consider
$\phi({\bf x})$ in a finite spatial volume  and impose periodic
boundary conditions. In such a case the momentum is discrete and
the evolution kernel (73) becomes a product of the evolution
kernels (12). The determinant of the operator
$-\partial_{s}^{2}+\nu^{2}$ with the Dirichlet boundary conditions
on $[0,t]$ appearing in eq.(69) can  be calculated and is equal to
$\nu^{-1}\sinh (\nu t)$. With $\nu\rightarrow\zeta \nu$ the
formula (69) still holds true while the calculation of the
determinant gives the result  $(\zeta\nu)^{-1}\sinh (\zeta\nu t)$.
Eq.(69) leads to the standard loop expansion with the Euclidean
propagator $G=iG^{E}$ (72) (which has an extra factor of $i$ in
comparison to the standard Euclidean field theory for the
$\tilde{V}$ interaction).   If $\tilde{V}^{\prime\prime}\leq
\mu^{2}$ in eq.(73) then the approximation
$\tilde{V}^{\prime\prime}= const$ may be applicable leading to the
expression (73) with
    $\nu=\sqrt{-\triangle + \mu^{2} -\tilde{V}^{\prime\prime}}$.

\section{Field theory on the lattice }
We already know from secs.1-7 how to define quantum mechanics of
an upside-down oscillator using various methods: Heisenberg
picture,  Schr\"odinger formulation, path integral or
stochastic representation. We repeat these methods in this section
in application to field theory. There can be two ways to view QFT
as a limit of quantum mechanics. The first method  assumes a
finite volume with periodic boundary conditions. Then, the momenta
form a discrete set which when cut to a finite number give a
quantum mechanical approximation. The second method is to set the
spatial coordinates of the field on a  lattice of a finite volume
with periodic boundary conditions. In both approaches the
transition to quantum field theory is straightforward at least for
free field theory. We discuss the second approach with
$(\nabla\phi)^{2} \rightarrow -(\nabla\phi)^{2}$ as discussed in
sec.8.

  We divide the Euclidean space $R^{d-1}$
into boxes of volume $\delta^{d-1}$. The scalar field $\phi$
depends on vertices of these boxes (called sites). We write
$\phi({\bf x})$ as $\phi(n_{1}\delta,.....,n_{d-1}\delta)$, where
$n_{j}$ are integers ( ${\bf n}\in Z^{d-1}$).
 We replace the derivative
in the Lagrangian by the lattice derivative
\begin{equation}
(\nabla_{j}^{\delta}\phi)({\bf
    n}\delta)=\delta^{-1}\Big(\phi(...,n_{j}\delta+\delta,....)-
\phi(.....,n_{j}\delta,...)\Big).
\end{equation}
The lattice Laplacian can be written as
\cite{simonpfi}\begin{equation} (\triangle_{\delta}\phi)({\bf
n}\delta)=\delta^{-2}\Big(\sum_{j} \phi({\bf n}\delta+{\bf
e}_{j}\delta)-2(d-1)\phi({\bf n}\delta)\Big),
\end{equation} where ${\bf e}_{j}$ is the unit vector in the $j$-th direction
and the sum is over the nearest neighbors of the point ${\bf
    n}\delta$  in all directions ${\bf e}_{j}$. The free field
Lagrangian is
\begin{equation}\begin{array}{l}
{\cal L}_{0}=\frac{1}{2}\sum_{\bf n}\delta^{d-1}(\frac{d\phi({\bf
n})}{dt})^{2}-\frac{1}{2}\delta^{d-1}\sum_{{\bf n}}\phi({\bf
    n}\delta)(\triangle_{\delta}\phi)({\bf
    n}\delta)+\frac{\mu^{2}}{2}\delta^{d-1}\sum_{{\bf n}}\phi({\bf
    n}\delta)^{2}\cr
\equiv \frac{1}{2}\sum_{\bf n}\delta^{d-1}\Big((\frac{d\phi({\bf
n})}{dt})^{2}+ ((\nu\phi)({\bf n}))^{2}\Big),
\end{array}
\end{equation}
where $\nu$ is a positive definite operator.

 The interaction has
the form
\begin{equation}
{\cal A }_{I}=\int ds\sum_{{\bf n}}\tilde{V}(\phi_{s}({\bf
n}\delta))\delta^{d-1}.
\end{equation}
The quantum field theory is defined  by the formal integral (we
set $\hbar=1$ in this section,${\cal A}={\cal A}_{0}+{\cal
A}_{I}=\int dt ({\cal L}_{0}+{\cal L}_{I}$))
\begin{equation}
d\mu_{\delta}=\prod_{{\bf n}}d\phi_{t}({\bf n}\delta)\exp(i{\cal
A}).
\end{equation}
We consider the Fourier transform of $\phi({\bf n}\delta)$
\begin{displaymath}
\tilde{\phi}_{\delta}({\bf k})=(2\pi)^{-\frac{d-1}{2}}\sum_{{\bf
        n}}\phi({\bf n}\delta)\exp(i{\bf kn}\delta).
\end{displaymath}Then, $\phi({\bf n}\delta) $ can be  expressed by
$\tilde{\phi}_{\delta}({\bf k})$ as
\begin{equation}
\phi({\bf
    n}\delta)=(2\pi)^{-\frac{d-1}{2}}\int_{-\frac{\pi}{\delta}}^{\frac{\pi}{\delta}}
d{\bf k} \exp( i{\bf kn}\delta)\tilde{\phi}_{\delta}({\bf k}),
\end{equation}where
the  ${\bf k}$-integration is over the cube
$-\frac{\pi}{\delta}\leq k_{j}\leq \frac{\pi}{\delta}$ .

If the lattice is infinite then the integral $\prod_{{\bf
        n}}d\phi_{s}({\bf n}\delta)\exp(i{\cal A}_{0}) $ should be understood
as the Gaussian integral on an infinite dimensional space of
sequences $\phi({\bf n}\delta)$. We can calculate from eq.(78) in
free field theory in an infinite volume (where the product
$\prod_{\bf n}d\phi_{t}({\bf n}\delta)$ is over all sites of the
lattice)
\begin{equation}\begin{array}{l}<\phi_{t}({\bf n}\delta)\phi_{t^{\prime}}({\bf
    n}^{\prime}\delta)>=Z^{-1}\int \prod_{\bf n}d\phi_{s}({\bf
    n}\delta)\exp(i{\cal A}_{0})\phi_{t}({\bf n}\delta)\phi_{t^{\prime}}({\bf
    n}^{\prime}\delta)
\cr=(2\pi)^{-d+1}\int_{-\frac{\pi}{\delta}}^{\frac{\pi}{\delta}}
d{\bf k} \exp( i({\bf kn}-{\bf
    kn}^{\prime})\delta)i(-\partial_{s}^{2}+\nu^{2}_{\delta})^{-1}(t,t^{\prime}),\end{array}
\end{equation}
where in eq.(80) $d\phi_{s}\exp(\frac{i}{2}\int
(\frac{d\phi_{s}}{ds})^{2})$ is expressed by the Wiener measure as
in sec.5 and
\begin{displaymath}Z=\int \prod_{\bf n}d\phi_{t}({\bf
    n}\delta)\exp(i{\cal A}_{0})
\end{displaymath}
\begin{equation}
\nu_{\delta}^{2}({\bf k})= (2d-2-2\sum_{j=1}^{d-1}\cos(\delta
k_{j}))\delta^{-2}+\mu^{2}
\end{equation}
and \begin{equation}
(-\partial_{t}^{2}+\nu_{\delta}^{2})^{-1}({\bf n}\delta,{\bf
n}^{\prime}\delta) =\Big((2\nu)^{-1}\exp(-\nu\vert
t-t^{\prime}\vert)\Big)({\bf n}\delta,{\bf n}^{\prime}\delta)
\end{equation}
It is understood that in eq.(80) we first consider a finite volume
cutoff and subsequently take the infinite volume limit . For the
quantum mechanics in a finite number of dimensions we can repeat
the calculations of one dimensional quantum mechanics with the
result (73).

The operator $\nu^{2}$ connects neighboring sites. So for a finite
number of $\phi({\bf n}\delta)$ we have a problem to define the
action of $\nu$ upon the last $\phi({\bf n}\delta)$. We can avoid
this difficulty for a finite lattice introducing periodic boundary
conditions on the lattice (then the last $\phi({\bf n}\delta)$ has
the first $\phi({\bf n}\delta)$ as a successor).

We consider  the Schr\"odinger equation
\begin{equation}i\hbar\partial_{t}\psi= \sum_{{\bf
n}}\Big(-\frac{\hbar^{2}}{2}\frac{\partial^{2}}{\partial\phi({\bf
n}\delta)\partial\phi({\bf n}\delta)}-\frac{1}{2}((\nu\phi)({\bf
n}\delta))^{2}\Big)\psi
\end{equation}
with the initial condition
\begin{equation}
\psi(\phi)=\exp(\frac{i}{2} \phi\nu\phi)\chi .\end{equation}

 We can repeat the formula
(28) expressing the time evolution on the periodic lattice by the
stochastic process solving the equation
\begin{equation}
d\phi_{t}({\bf n}\delta)=-(\nu\phi)({\bf n}\delta)dt+\sigma
dW({\bf n}\delta)  .
\end{equation}
The solution is
\begin{equation}
\phi_{t}(\phi,{\bf n}\delta)=(\exp(-\nu t)\phi)({\bf n}\delta) +
\sigma\int_{0}^{t}\Big(\exp(-\nu(t-s))dW_{s}\Big)({\bf n}\delta) ,
\end{equation}
 where the Brownian motions $W_{s}({\bf n}\delta)$ are defined
as the mean zero Gaussian processes with $W_{0}=0$ and
\begin{equation}
E[W_{t}({\bf n}\delta)W_{s}({\bf
n}^{\prime}\delta)]=min(t,s)\delta({\bf n},{\bf n}^{\prime})
\end{equation} with the Kronecker $\delta-$ function on the rhs of
eq.(87).     In such a case eq.(18) still holds true in the form
\begin{equation} \psi_{t}= \exp(-\frac{t}{2}Tr\nu
)\exp(\frac{i}{2} \phi\nu\phi)\chi_{t}  .
\end{equation}
We can express the solution of eq.(83) on a finite lattice in the
form
\begin{displaymath}\begin{array}{l}
\psi_{t}(\phi)=\exp(-\frac{t}{2}Tr\nu+\frac{i}{2}\phi\nu\phi)E[\chi(\phi_{t}(\phi))]\cr=
\int
K(t;\phi,\phi^{\prime})\exp(\frac{i}{2}\phi^{\prime}\nu\phi^{\prime})\chi(\phi^{\prime})d\phi^{\prime}
\end{array}\end{displaymath}
The lhs of this equation (without the $\exp(-\frac{t}{2}Tr\nu) $
term ) has a meaning in the continuum with an infinite volume,
whereas we would have some difficulties defining the rhs because
the formal Lebesgue measure $d\phi$ has no meaning  in an infinite
number of dimensions.

\section{Feynman integral in  QFT of exponential interactions}
We define the solution of the continuum version of the
Schr\"odinger equation (83) by the formula (omitting the term
$\exp(-\frac{t}{2}Tr \nu)$ as the vacuum energy renormalization)
\begin{equation}\begin{array}{l}
\psi_{t}(\phi)=\exp(\frac{i}{2\hbar}\phi\nu\phi)E[\chi(\phi_{t}(\phi))],
\end{array}\end{equation}with (see \cite{daletski}\cite{kuo} for
stochastic equations in infinite number of dimensions)
\begin{equation}
\phi_{t}(t_{0},\phi)=\exp(-\nu
(t-t_{0}))\phi+\sigma\int_{t_{0}}^{t}\exp(-\nu(t-s))dW_{s},
\end{equation}
where $\nu=\sqrt{-\triangle+\mu^{2}}$ and $W_{t}({\bf x})$ is the
Gaussian process with the covariance
\begin{equation}
E[W_{t}({\bf x})W_{s}({\bf x}^{\prime})]=min(t,s)\delta({\bf
x}-{\bf x}^{\prime})
\end{equation}
We could also define    the solution with the initial value at $t_{0}=-\infty$
\begin{equation}
\phi_{t}=\lim_{t_{0}\rightarrow
-\infty}\phi(t_{0},\phi)=\sigma\int_{-\infty}^{t}\exp(-\nu(t-s))dW_{s}  .
\end{equation}
From eq.(90) we obtain the correlation function (30) in field
theory as
\begin{equation}\begin{array}{l}
E[\phi_{t}(\phi,{\bf y})\phi_{s}(\phi,{\bf x})]=
(\exp(-(t-t_{0})\nu)\phi)({\bf y})(\exp(-(s-t_{0})\nu)\phi)({\bf
x})\cr+\Big(\frac{1}{2\nu}\exp(-\nu(t+s-2t_{0}))\Big)({\bf x},{\bf
y})+G(t,{\bf y};s,{\bf x}), \end{array}\end{equation} where
\begin{equation}\begin{array}{l}
G(t,{\bf y};s,{\bf x})=i(-\partial_{0}^{2}-\triangle
+\mu^{2})^{-1}\cr=\frac{i}{2}\Big(\nu^{-1}\exp(-\nu\vert t-s\vert)\Big)({\bf
x},{\bf y})=iG^{E}(t,{\bf y};s,{\bf x}) \end{array}\end{equation}
The first two terms on the rhs  of eq.(93) are regular for $t+s>2t_{0}$. In
the limit $t_{0}\rightarrow -\infty$ the expectation value (93)
tends to $iG^{E}$ where $G^{E}$ is the Euclidean  correlation
function for (quantum) free fields.

For a study of an analytic continuation ${\bf x}\rightarrow i{\bf
x}$ we change $\nu\rightarrow \zeta\nu$ where $\zeta
=\alpha+i\beta$ . Then, eq.(90) reads  (we could also consider a
modification of eq.(95) for a dynamical $\zeta$ depending on time)
\begin{equation}
\phi^{\zeta}_{t}(t_{0},\phi)=\exp(-\zeta\nu
(t-t_{0}))\phi+\sigma\int_{t_{0}}^{t}\exp(-\zeta\nu(t-s))dW_{s}
\end{equation}
and the last term in eq.(93) is
\begin{equation}
G^{\zeta}(t,{\bf y};s,{\bf
x})=\frac{i}{2}\Big(\zeta^{-1}\nu^{-1}\exp(-\nu\zeta\vert
t-s\vert)\Big)({\bf x},{\bf y})
\end{equation}
$\exp(-\zeta\nu t)$ defines  an analytic semigroup \cite{nelson}
which is well-defined for any real $\beta$ and $\alpha\geq 0$. In
particular, eq.(96) for $\zeta=i$ defines the Feynman propagator
in the Minkowski space-time.
 In
order to define an interaction we need to regularize the field
(95). This can be done either by a replacement of the delta
function on the rhs of eq.(91) by a regular function or replacing
$\nu$ by $\nu_{\epsilon}=\nu(1+\epsilon k^{4})$ (in momentum
space, in coordinate space $k^{4}\rightarrow \triangle^{2}$). In
such a case the field correlation function in eq.(93) is a regular
function (because the kernel of $\nu_{\epsilon}^{-1}$ is a regular
function in $d\leq 4$).

We   derive a solution of the Schr\"odinger equation for the
Hamiltonian $H=H_{0}+\tilde{V}$ in the form
\begin{equation}\begin{array}{l}
\psi_{t}(\phi)= \exp(\frac{i}{2\hbar}\phi\nu\phi) \cr
E\Big[\exp\Big(-\frac{i}{\hbar}\int_{t_{0}}^{t}
\tilde{V}(\phi_{s}(\phi,{\bf x}))d{\bf
x}ds\Big)\chi(\phi_{t}(\phi))\Big].
\end{array}\end{equation}or
with a scaled initial condition\begin{equation}\begin{array}{l}
\psi_{t}(\sigma\phi)= \exp(-\frac{1}{2}\phi\nu\phi) \cr
E\Big[\exp\Big(-\frac{i}{\hbar}\int_{t_{0}}^{t}
V(\phi_{s}(\sigma\phi,{\bf x}))d{\bf
x}ds\Big)\chi(\phi_{t}(\sigma\phi))\Big].
\end{array}\end{equation}
We show that for some potentials the expressions (97)-(98) defined
first for regularized fields have a limit when
$\epsilon\rightarrow 0$. For exponential interactions the
existence of this limit can be shown in a convergent perturbation
series and for some polynomial interactions in a non-perturbative
way.

First, let us consider the exponential interaction
\begin{equation}
\tilde{V}(\phi)=g\int d{\bf x}\int d\mu(a)\exp(ia\alpha\phi({\bf
x}))
\end{equation}

 The potential
in eq.(97) enters as an exponential of the expression
\begin{equation}\begin{array}{l}
\int dsd{\bf x}\tilde{V}(\phi_{s}^{\epsilon}(\phi,{\bf x})) =g\int
dsd{\bf x}d\mu(a)\exp\Big(ia\alpha\Big((\exp(-\nu_{\epsilon}
(s-t_{0}))\phi)({\bf x})\cr+\sigma
\int_{t_{0}}^{s}(\exp(-\nu_{\epsilon}(s-\tau))dW_{\tau})({\bf
x})\Big)\Big)  .
\end{array}\end{equation} We need a normal ordering of this expression
\begin{equation}\begin{array}{l}
\int dsd{\bf x}d\mu(a):V(\phi_{s}^{\epsilon}({\bf x})): \cr=g\int
dsd{\bf x}d\mu(a)\exp\Big(ia\alpha\Big(\exp(-\nu_{\epsilon}
(s-t_{0}))\phi)+ia\alpha\Big(\sigma
\int_{t_{0}}^{s}\exp(-\nu_{\epsilon}(s-\tau)dW_{\tau}\Big)\Big)\cr
\Big(E\Big[\exp\Big(\sigma ia\alpha
\int_{t_{0}}^{s}\exp(-\nu_{\epsilon}(s-\tau)dW_{\tau}\Big)\Big]\Big)^{-1}.\end{array}
\end{equation}
We show that correlation functions of the normal ordered
exponential interaction considered as generalized functions have a
 limit when $\epsilon\rightarrow 0$. The n-point functions of
normal ordered exponentials have the form (without the
$\phi$-term)
\begin{equation}
\begin{array}{l}
\int_{\Omega}d{\bf x}_{1}.....d{\bf
x}_{n}ds_{1}...ds_{n}d\mu(a_{1})...d\mu(a_{n})\prod_{j\neq k}\cr
\exp\Big(-\frac{1}{2}\sigma^{2}\alpha^{2}a_{j}a_{k}\int_{t_{0}}^{\min(s_{j},s_{k})}
\Big(\exp(-(s_{j}+s_{k}))\nu_{\epsilon})\exp(2\tau\nu_{e})\Big)({\bf
x}_{j},{\bf x}_{k})d\tau\Big)  .
\end{array}
\end{equation}The condition $j\neq k$ comes from the normal
ordering. These correlation functions will appear in the expansion
of the Feynman-Kac formula (97) in powers of $g$ as in eq.(49).
The integral over $\tau$ in (102) is expressed by ( and the terms
in eq.(93) which are regular even in the limit $
\epsilon\rightarrow 0$ and vanish when $t_{0}\rightarrow -\infty$
)
\begin{equation}
G^{E}_{\epsilon}(s_{j},s_{k})=\frac{1}{2}\Big(\nu_{\epsilon}^{-1}\exp(-\nu_{\epsilon}
\vert s_{j}-s_{k}\vert )\Big)({\bf x},{\bf x}^{\prime})
\end{equation}
When $\epsilon \rightarrow 0$ the Green  function (103) becomes
singular at coinciding points on a set of Lebesgue measure $0$. As
$\sigma^{2}\alpha^{2}=\pm i\hbar$ the product of exponentials
(102) is a bounded function. The Lebesgue integral over a bounded
domain $\Omega$ satisfies
\begin{equation} \vert \int_{\Omega} f d{\bf x}\vert \leq \sup
\vert f\vert \vert\Omega\vert \end{equation} where
$\vert\Omega\vert$ is the volume. For  $ f=\exp(iu)$, where $u$ is
real we have
\begin{equation}
\vert \int_{\Omega} f d{\bf x}\vert \leq \vert\Omega\vert.
\end{equation}
It follows from  eq.(105) and the Lebesgue dominated convergence
theorem that the limit $\epsilon\rightarrow 0$ of the correlation
functions (102) exists in any dimension. In the limit
$\epsilon\rightarrow 0$ the set of singular points of the terms in
the exponential (102) is of measure zero. Hence, the limiting
correlation functions are well-defined as distributions in any
dimension ( with our choice of regularization in $d\leq 4$).
                                              We can
take the limit $\epsilon \rightarrow 0$ at each order of the
perturbation series in $g$. From the bound (105) it follows that
the perturbation series is convergent.     On the basis of similar
estimates as in eq.(52) we obtain the perturbative  Feynman-Kac
formula for
 non-polynomial interactions which have been of interest for
field theory for a long time \cite{salam}. The formula applies for
an inverted sign of the metric. It is unclear how it could be
continued analytically either to the quantum field theory in
Euclidean or Minkowski space-time in more than two dimensions. The
exponential interaction in $d=2$ appears in Polyakov string theory
\cite{polyakov}\cite{tischner}. The four-dimensional exponential
interaction is applied in the Starobinsky model of inflation
\cite{starobinsky} resulting from an interaction of a scalar field
with Einstein gravity.

 There remains the question of whether we can define the
matrix elements $(\psi^{\prime},\psi_{t})=
  (\psi^{\prime},U_{t}\psi)$ of the evolution operator in a Hilbert space. First of all we
  should mention that there is no distinguished Hilbert space  for
  the upside-down oscillator. In the standard formulation of the Feynman-Kac
  formula in QFT \cite{feldman}
we are interested in a computation of matrix elements between
particle states. This means that in the $L^{2}(d\mu_{A})$
representation of the Fock space, $d\mu_{A}$ is the Gauss measure
with the covariance $A^{-1}=(-\triangle +m^{2})^{-\frac{1}{2}}$,
where $m$ is the particle mass. With the upside-down (infinite
dimensional) oscillator there is no Fock space to begin with. We
may be interested in the calculation of matrix elements of the
evolution operator between other states. In any case the choice of
$A=\nu=\sqrt{-\triangle +\mu^{2}}$ will lead to ultraviolet
divergences   in a calculation of matrix elements of the evolution
operator in more than two dimensions as will be seen from the
calculations below.
   Let us consider
  the $\phi$-term   in the matrix elements $(\psi^{\prime},U_{t}\psi)$ resulting from the perturbative Feynman-Kac
  formula integrated
   with respect to a Gaussian measure
   \begin{equation}
   d\mu_{A}=d\phi\exp(-\frac{1}{2}(\phi, A\phi))
   \end{equation} with a certain operator $A$.
  A function of $\phi$, coming from the calculation of the expectation value in the the Feynman-Kac factor
  in the formula (97),
   will in general be a distribution (if $\phi$ is a distribution from the support of the measure $d\mu_{A}$).
   Then, the integral over $\phi$ can give either zero or
   infinity.
   Let us consider  the exponential interaction
   (99) with  $\alpha=i$ . Then, the $\phi$ term separates from the expectation
   value $E[..]$ as
   \begin{equation}
   I(\phi)=\exp\Big(-g\int_{0}^{t}dsd{\bf x}\int d\mu(a)\exp(ia\alpha (e^{-\nu s}\phi)({\bf
   x})\Big)
   \end{equation}
    If $\alpha =i$, $g\geq 0$  and $ d\mu(a) $ is a non-negative measure then we have a non-positive function in the
    exponential (107).
    Hence, $\int d\mu_{A}I(\phi)\leq 1$, From the Jensen inequality
    \begin{equation}\begin{array}{l}
    \int d\mu_{A}I(\phi)\geq\exp\Big(-g\int_{0}^{t}dsd{\bf x}\int d\mu(a)\int d\mu_{A}(\phi)\exp(-a (e^{-\nu s}\phi)({\bf
    x})\Big),
   \end{array} \end{equation} where
      \begin{equation}
      \int d\mu_{A}\exp(-a (e^{-\nu s}\phi)({\bf x}))  =\exp\Big(
      \frac{1}{2}a^{2}
      (\exp(-\nu s)A^{-1}\exp(-\nu s))({\bf x},{\bf x})\Big)
      \end{equation}
      With $A=1$ we have  (if  $\nu=\sqrt{-\triangle}$ )
       \begin{displaymath}
       (\exp(-\nu s)A^{-1}\exp(-\nu s)({\bf x},{\bf x})\Big)
       =s^{-1}
       \end{displaymath}
       in space-time dimension  $d=2 $ and  \begin{displaymath}
 (\exp(-\nu s)A^{-1}\exp(-\nu s)({\bf x},{\bf x})\Big)
  =\frac{1}{8}\pi s^{-3}
  \end{displaymath}
  in $d=4$. Hence, the $s$-integral is infinite in eq.(108). The lower bound in eq.(108)
  is trivial. In order to get a non-zero lower bound we need a
  differential
  operator $A$. In d=2 it is sufficient to take $A=\nu^{\delta}$
  and in $d=4$ ,    $A=\nu^{4+\delta}$ with $\delta>0$.
  We can study the $\phi$-integral of the expression (107) in a perturbation expansion
  in $g$. The first order in $g$ is the exponential of eq.(108) which we have discussed already.
  At the order $g^{2}$
 the integral over $d\mu_{A}$ of (107)is
\begin{displaymath}\begin{array}{l}
 \int d\mu_{A}(\phi) d\mu(a_{1})d\mu(a_{2}) \int_{0}^{t}ds_{1}\int_{0}^{t}ds_{2}d{\bf x}_{1}  d{\bf x}_{2}
\cr\exp(ia_{1}\alpha (\exp(-\nu s_{1})\phi)({\bf x}_{1}))
\exp(ia_{2}\alpha (\exp(-\nu s_{2})\phi)({\bf x}_{2}))
\end{array}\end{displaymath}
It can be  expressed by
  \begin{equation}
  \exp\Big(-\frac{1}{2} a_{1}a_{2}\alpha^{2}
  \Big(\exp(-\nu s_{1})A^{-1}\exp(-\nu s_{2} )\Big)({\bf x}_{1},{\bf x}_{2})  \Big)
  \end{equation}
  With $A=1$  and $\nu=\sqrt{-\triangle}$ we have  in $d=2$
     \begin{equation}
     (\exp(-\nu s_{1})A^{-1}\exp(-\nu s_{2} )({\bf x}_{1},{\bf x}_{2})  \Big)
     =2(s_{1}+s_{2})((s_{1}+s_{2})^{2}+(x_{1}-x_{2})^{2})^{-1} .
     \end{equation}
     and in $d=4$
  \begin{equation}
  (\exp(-\nu s_{1}))A^{-1}\exp(-\nu s_{2} )({\bf x}_{1},{\bf x}_{2})
  =\pi (s_{1}+s_{2}) \Big( (s_{1}+s_{2})^{2}+({\bf x}_{1}-{\bf x}_{2})^{2}\Big)^{-2}
  .
  \end{equation} The expressions (111)-(112) appearing in the
  exponential (110) will be singular if $A$=1 when $s_{1}=s_{2}=0$ and ${\bf x}_{1}={\bf x}_{2}$. We need $ A
  =\nu^{1+\delta}$ in $d=2$ and $A=\nu^{d-1+\delta}$ ($\delta>0$ )in $d$
  dimensions if these terms are to be finite. It can be seen that
  this condition for $A$ is sufficient for finite $d\mu_{A}(\phi)$
  correlations at any order  of $g$.

Although  the correlation functions in the perturbative expansion
(102) are well-defined distributions and the perturbative
expansion is convergent the
 potential in the Feynman-Kac formula
\begin{equation} \int_{0}^{t}dsd{\bf x}
:V(\phi_{s}(\phi,{\bf x})):=g\int_{0}^{t}dsd{\bf x} :\exp(ia\phi_{s}(\phi,{\bf x})):
\end{equation}
is a  square integrable random variable only in two space-time
dimension and  if $\hbar a^{2}<2\pi$. In fact,
\begin{equation}\begin{array}{l}
E\Big[\int_{0}^{t}dsdx :V(\phi_{s}(\phi;x)):\int_{0}^{t}dsdx
(:V(\phi_{s}(\phi;x)):)^{*}]=g^{2}\int_{0}^{t}dsdx\cr\int_{0}^{t}ds^{\prime}dx^{\prime}
E[:\exp(ia\phi_{s}(\phi;x))::\exp(-ia\phi_{s^{\prime}}^{*}(\phi;x^{\prime})):]
\end{array}\end{equation}
The field $\phi^{*}=\exp(-\nu s)\phi +\sigma^{*}b$ (where $b$ is a
real random variable) depends on $\sigma^{*}$. As a consequence in
the expectation value (114) instead of the term
$\sigma^{2}=i\hbar$ (as in eq.(102)) we shall have
$\sigma^{*}\sigma=\hbar$ . This leads to a singular integral
\begin{equation}
\int dsds^{\prime}dxdx^{\prime}\exp\Big(-a^{2}\hbar\int
\frac{1}{2\nu}\exp(-\nu \vert
s-s^{\prime}\vert)(x,x^{\prime})\Big)
\end{equation}
which  is finite only in one space dimension and  if $\hbar
a^{2}<2\pi$ ( for exponential models in two space-time dimensions
see \cite{ah}\cite{gallavotti}\cite{frohlich}).

It follows that for an exponential  and the trigonometric
interactions after the normal ordering the limit
$\epsilon\rightarrow 0$  at each order of the perturbation series
in $g$ exists. The perturbation series in $g$ is convergent and
the Feynman-Kac formula (97) solves the Schr\"odinger equation
with the potential (99). Now the question can be raised how to
continue this theory to the quantum field theory in the Minkowski
space-time or to Euclidean field theory. In the latter case we
could perform an analytic continuation from Euclidean field theory
to the quantum (Wightman) field theory by means of the
Osterwalsder-Schrader reconstruction \cite{os}. It can be seen
from eq.(96) that the analytic continuation ${\bf
k}^{2}\rightarrow -{\bf k}^{2}$ and $\mu^{2}\rightarrow -\mu^{2}$
($\nu_{\epsilon}\rightarrow i\nu_{\epsilon}$) has the effect  that
the exponential (102) is not a pure phase anymore (because $i$ in
$\sigma^{2} $ is cancelled by $\frac{1}{i\nu_{\epsilon}}$ and
there will be $i$ in $\exp(-i\nu\vert s-s^{\prime}\vert))$. After
the analytic continuation the correlations (102) will have a
non-integrable singularity in $d>2$. In one spatial dimension the
expression (102) behaves at short distances as
\begin{equation}
\prod_{j\neq k}\exp(-\hbar a^{2}\frac{1}{2\pi}\ln(
(x_{j}-x_{k})^{2}-(s_{j}-s_{k})^{2})
\end{equation}
Hence, for $\hbar a^{2}< 2\pi$ each term of the perturbation
series in $g$ is integrable. In the Euclidean domain the
perturbation series in the trigonometric model is convergent
\cite{french}  . Then, the analytic continuation to the quantum
field theory in the two-dimensional Minkowski space-time can be
achieved by means of the Osterwalder-Schrader reconstruction
theorem \cite{os}.

\section{QFT of polynomial interactions}

  In this section  discuss as another example the polynomial interactions in two dimensions of the form
 \begin{equation}V(\phi)=g:\phi_{s}^{6}:
 \end{equation} (in this example we could work without the scaling $\sigma$) and
\begin{equation}V(\phi)=g:\phi_{s}^{4}:.
 \end{equation} in any dimension.
The field is expressed by the Ornstein-Uhlenbeck process $\xi_{s}$
as $\phi_{s}(\phi)=\sigma(\exp(-\nu s)\phi+ \xi_{s}$). At short
distances $\xi_{s}({\bf x})$ has the same singularities as
 the free scalar Euclidean field (as seen from eq.(93)).

 For the normal ordering of $\sigma \xi$ we have
\begin{equation}
-i:(\sigma\xi)^{6}=-i\sigma^{6}:\xi^{6}:=-\hbar^{3}:\xi^{6}:
\end{equation}  (by the normal orderings in eqs.(117)-(119) we mean the normal  ordering of $\xi_{s}).$
 Hence, there is the same  semi-boundedness of
$\exp(-i:(\sigma\xi)^{6}))$ as in Euclidean $\phi^{6}$. For
$\phi^{4}$ in eq.(98)

\begin{equation} -i:(\sigma\xi)^{4}=-i\sigma^{4}:\xi^{4}:=i\hbar^{2}:\xi^{4}:
\end{equation}
Hence,  we have an oscillatory function in the exponential of the
Feynman-Kac formula (98).

  Let us consider $\phi^{6}$ in two dimensions first.
We can show (following \cite{grsIHP}) that $V(\phi(s,x))$ has  well-defined locally
integrable correlation  functions  in $d=2$
\begin{equation}
E[V(\phi(s_{1},x_{1})).....V(\phi(s_{n},x_{n})]
\end{equation}
Moreover, for any $n$
\begin{equation}
E\Big[\Big\vert \int_{0}^{L} dx\int_{0}^{t}ds
V(\phi_{s}(\phi,x))\Big\vert^{2n}\Big]<\infty
\end{equation}
 We introduce the volume cutoff ( $x\in [0,L]$) and the ultraviolet cutoff $\epsilon$ such that
 $\nu_{\epsilon}=\kappa_{\epsilon}(k)\nu$ with $\kappa_{\epsilon}(k)=1+\epsilon
 k^{4}$.
 The regularized fields $\phi^{\epsilon}$ (90)
 are solutions of the stochastic equation  with the regularized
 $\nu_{\epsilon}$.
 Then, we have (we
treat the initial condition $\phi$ as a regular external field)
 \begin{equation} \Big\vert
 \exp\Big(-\frac{i}{\hbar}\int_{0}^{t}ds\int_{0}^{L}dx
                             : V(\phi^{\epsilon}_{s}(\sigma\phi)):\Big)\Big\vert
 \leq
 R(\phi)\exp\Big(tLK(\ln
 \frac{1}{\epsilon})^{3}\Big)\end{equation} with a certain constant $K$.
  This estimate  follows from an estimate of the lower bound of the sixth order polynomial whose coefficients
  at the $2n$-th order term behave     \index{polynomial interaction}
  as $<\phi_{\epsilon}^{2}(x)>^{3-n}$ where $n=1,2$.
  In order to prove the limit $\epsilon\rightarrow 0$ we can
  apply the Duhamel expansion of refs.\cite{gj}\cite{grsIHP}
  This expansion arises from the identity
 (see \cite{grsIHP}, sec.VII.4)
\begin{equation}
\exp(-U)=\exp(-U_{\epsilon_{1}})-\int_{0}^{1}\exp(-\tau_{1}U)\delta U_{\epsilon_{1}}\exp(-(1-\tau_{1})U_{\epsilon_{1}})d\tau_{1}
\end{equation}
where $U_{\epsilon}=\int \tilde{V}(\phi_{\epsilon})$ and $\delta
U_{\epsilon}=U-U_{\epsilon}$.  The iteration of eq.(124) gives
\begin{equation}
E\Big[\exp(-U)\Big]=E\Big[\sum_{n=0}^{\infty}
(-1)^{n}\int...\int\prod
d\tau_{1}..d\tau_{n}\prod_{j=1}^{n+1}\exp(-\delta
s_{j}U_{\epsilon_{j}})\prod_{k=1}^{n}\delta U_{\epsilon_{k}}\Big]
\end{equation}
where $s_{0}=1$,$s_{n+1}=0$, $\delta s_{j}=s_{j-1}-s_{j}$ and
$s_{0}\geq s_{1}\geq......\geq s_{n+1}$.

 The method of the Duhamel expansion is applying the estimate
 (for natural $p\geq 1$)
 \begin{equation}
 E\Big[ \Big\vert     : V(\phi^{\epsilon}_{s}(\sigma\phi)):-: V(\phi_{s}(\sigma\phi)):
 \Big\vert ^{p} \Big]  < C^{p}(\phi) \epsilon^{\beta p}p^{3p}
 \end{equation}  with certain function $C(\phi)$ of the background field $\phi$ and
 a positive constant $\beta$. Then, in ref. \cite{grsIHP}   a special choice of  cutoffs
 $\epsilon_{j}=\exp(-j^{\frac{1}{3}})$ is applied so that the Feynman-Kac
 factor in eq.(123) is bounded by $\exp (jtLK)$ for the $j$-th
 cutoff.
 In such a case  the exponentially growing terms
 $\exp (jtLK)$ in eq.(125)
 are compensated by   the multipliers $ C(\phi)^{p}\exp(-\beta p j)p^{3p}$ of eq.(126) in the  Duhamel expansion
 (125). As a result
 the  series (125) is convergent proving
 that $E[\exp(-U)]$ is finite.

In the same way we can show that the limit of the  Feynman
 integral\begin{displaymath} \lim_{\epsilon\rightarrow
 0}E\Big[\exp\Big(-\frac{i}{\hbar}\int_{0}^{t}ds\int_{0}^{L}dx :V(\phi^{\epsilon}_{s}(\sigma\phi)):\Big)\chi(\sigma\phi_{t})\Big]
 \end{displaymath}
 exists for a  set of initial wave functions $\chi(\phi)$ of the form

$\chi(\phi)=\sum_{n}c_{n}\exp((\phi,f_{n}))$.

Next, we consider  $\phi^{4} $ model for the analytically
continued wave function $\psi_{t}(\sigma\phi) $. An estimate of
eq.(98) is simpler as the upper bound (123) is replaced by 1. In
any dimension the generating functional (where $u$ is a real
function and $\Omega$ is a bounded domain in $R^{d-1}$)
\begin{equation}\begin{array}{l}
Z_{\epsilon}[u]=E\Big[\exp\Big(-\frac{i}{\hbar}\int_{0}^{t}ds\int_{\Omega}d{\bf
x}g:\phi^{\epsilon}_{s}(\sigma\phi,{\bf x}))^{4}:\Big)
\cr\times\exp\Big(\int dsd{\bf x} u(s,{\bf
x})\phi^{\epsilon}_{s}(\sigma\phi,{\bf x})\Big)\Big].
\end{array}
\end{equation}has a limit when $\epsilon\rightarrow 0$ as
\begin{equation}\begin{array}{l}
\Big\vert\exp\Big(-\frac{i}{\hbar}\int_{0}^{t}ds\int_{\Omega}d{\bf
x} g:\phi^{\epsilon}_{s}(\sigma\phi,{\bf x}))^{4}:\Big) \exp(\int
ds d{\bf x}u(s,{\bf x})\phi^{\epsilon}_{s}(\sigma\phi,{\bf
x}))\Big\vert \cr \leq\exp\Big(\int dsd{\bf x} u(s,{\bf
x})\Big((\exp(-\nu s)\phi)({\bf x})+
\frac{1}{\sqrt{2}}\xi^{\epsilon}_{s}({\bf x})\Big) \Big).
\end{array}\end{equation}
Therefore, the limit $lim_{\epsilon\rightarrow 0} Z_{\epsilon}(u)$
exists on the basis of the Lebesgue convergence theorem and
satisfies the bound (from eq.(93))
\begin{equation} \begin{array}{l}
\vert Z[u]\vert\leq \exp\Big(\int ds d{\bf x}u(s,{\bf
x})(\exp(-\nu s)\phi)({\bf x})\cr+

\frac{1}{4}\int dsds^{\prime} d{\bf x} d{\bf x}^{\prime}
u(s^{\prime},{\bf x}^{\prime})G^{E}(s,{\bf x};s^{\prime},{\bf
x}^{\prime})u(s,{\bf x})

\cr +\frac{1}{8}\int ds ds^{\prime}d{\bf x} d{\bf x}^{\prime}
 u(s^{\prime},{\bf x}^{\prime})\Big(\nu^{-1}\exp(-\nu(s+s^{\prime}))\Big)({\bf x},{\bf x}^{\prime})u(s,{\bf x})
\Big)\end{array}\end{equation} The continuation to Minkowski
space-time $\nu\rightarrow i\nu$ or to the Euclidean (imaginary
time) theory $t\rightarrow it$ seems possible only in
two-dimensions when we can repeat the argument (123)-(126) of
\cite{gj}\cite{grsIHP} as we did in the $\phi^{6}$ model.

In the estimate of the matrix elements \begin{equation}
(\psi^{\prime}, U_{t}\psi)=\int
d\phi\psi^{\prime}(\phi)\psi_{t}(\phi)
\end{equation}with $\psi_{t}(\sigma\phi)$  calculated from the Feynman integral
 we would need to apply the Cauchy formula
of the form (58). The measure (106) does not have the
$\sigma$-continuation. Hence, a calculation of the integral (130)
between particle states  does not seem possible( the Fock space
can be represented as $L^{2}(d\mu_{A})$ with
$A=\sqrt{-\triangle+m^{2}}$, where $m$ is the particle mass ). As
possible states for an analytic continuation of the integral (130)
we may consider states $\psi^{\prime}$ with the phase factor
$\exp(\frac{i}{2}\phi A\phi)$ which after the sigma continuation
becomes $\exp(-\frac{\hbar}{2}\phi A\phi)$. Assuming $A\simeq
\nu^{n}$ with a sufficiently large $n$ the support of the Gaussian
measure $d\mu_{A}$ is on continuous functions. In such a case the
first term on the rhs of eq.(58) may be well-defined. There
remains the problem whether the second term on the rhs can be
defined in an infinite number of dimensions. It seems however that
if we resign of calculating matrix elements between particle
states then QFT based on the Feynman formula can make sense in
higher dimensions at least with an inverted sign of the signature.
It is rewarding that the analytic continuation ${\bf x}\rightarrow
i{\bf x}$ and $\phi \rightarrow \sigma\phi$ in the Feynman-Kac
formulas (97)-(98) gives a substantial simplification in the proof
of the existence of the solution of the Schr\"odinger equation via
the Feynman-Kac formula. We hope that in more than two dimensions
there may be another analytic continuation method leading to the
quantum field theory in Minkowski space-time or that the approach
presented in this paper will be useful in scalar field theories
coupled to quantum gravity where the inversion of the signature
$(\nabla \phi)^{2} \rightarrow -(\nabla\phi)^{2}$ may appear as a
physical phenomenon.

\section{Summary and outlook}
We have shown that a rigorous version (in real time) of the
Feynman-Kac formula for analytic perturbations of the upside-down
oscillator can be extended to QFT with an inverted signature
($(\nabla\phi)^{2}\rightarrow -(\nabla\phi)^{2}$).
 $\psi_{t}(\phi) $ in
eq.(97) for exponential and trigonometric interactions and
$\psi_{t}(\sigma\phi) $ for the polynomial interaction $\phi^{4n}$
exist in higher dimensions as solutions of the Schr\"odinger
equation. There is still the problem whether this solution defines
a unitary evolution in a certain physical Hilbert space and
whether we could make the analytic continuation
$\sqrt{-\triangle}\rightarrow i\sqrt{-\triangle}$ in the resulting
final theory in order to obtain a Lorentz invariant model. The
technical reason for the simplification, in comparison to the
operator method in Minkowski space-time and the functional
integration techniques  in Euclidean field theory, is that the
inverted signature leads to oscillatory integrals in probabilistic
correlation functions. As a consequence the oscillatory integrals
can be bounded by a constant. The question remains open of whether
the resulting theory in higher dimensions is non-trivial and
whether the signature can be inverted back to the Minkowski
space-time. The latter procedure may be unnecessary in quantum
gravity when the signature can really change sign in some domains
of the space-time. In two-dimensions we can return to the
Euclidean formulation of QFT reversing the signature
$\sqrt{-\triangle}\rightarrow i\sqrt{-\triangle}$. Then, applying
the well-established results we can construct QFT in Minkowski
space-time by means the Osterwalder-Schrader reconstruction. We
hope that the inverted signature technique may lead to some new
methods for a construction of QFT in higher dimensions.

\section{Appendix:Sojourn time in  $ x^{6}$ potential}
The solution of the Schr\"odinger equation in $\tilde{V}=gx^{6}$
potential reads \begin{equation}\begin{array}{l}
\psi_{t}(x)=\exp(-\frac{\nu}{2}t)\exp(\frac{i\nu}{2\hbar}x^{2})
\cr E\Big[\exp\Big(-\frac{i}{\hbar}\int_{0}^{t}(\exp(-\nu
s)x+\sigma b_{s})^{6}ds \Big)\chi((\exp(-\nu t)x+\sigma
b_{t})\Big],
\end{array}\end{equation}where $b_{s}=\xi_{s}(0)$ is the
Ornstein-Uhlenbeck process (39). In order to estimate the sojourn
time we can consider the probability density $\vert\psi_{t}\vert^{2}$  with
\begin{equation}
\chi(x)=(2L)^{-\frac{1}{2}}R\exp(-\frac{x^{6}}{2L^{6}}) .
\end{equation} The numerical constant $R$ ($\simeq 1$) is chosen such
that $\int dx\vert \chi\vert^{2}=1$ so that  the probability
$\int_{-L}^{L}dx\vert\psi_{0}(x)\vert^{2}$ to find a particle initially in
the interval $[-L,L]$ is approximately equal to 1. We calculate
\begin{equation}\begin{array}{l}
 \int_{-L}^{L}dx \vert \psi_{t}(x)\vert^{2}= \int_{-L}^{L}dx (2L)^{-1}R^{2}
  \exp(-\nu t)E[\exp(F+F^{*})]\cr
  =    \int_{-L}^{L}dx (2L)^{-1}R^{2}\exp\Big(E[F+F^{*}]+
  \frac{1}{2}\Big(E[(F+F^{*})^{2}]-E[F+F^{*}]^{2}\Big) + ....\Big)
  .
   \end{array}
\end{equation}
In eq.(133) by $F$ we denote the exponential in eq.(131) and we applied the cumulant expansion for a calculation of an
expectation value of an exponential function.
 We restrict ourselves to the approximation
\begin{equation}
E[\exp (F+F^{*})]\simeq \exp(E[F+F^{*}]) .
\end{equation}
We can calculate the rhs of eq.(134) exactly using
\begin{equation}
E[b_{s}^{2}]=\frac{1}{2\nu}(1-\exp(-2\nu s))
\end{equation}
and
\begin{equation}
E[b_{s}^{6}]=15\Big(\frac{1}{2\nu}(1-\exp(-2\nu s))\Big)^{3} .
\end{equation}
The expression  (134) is still complicated so we write down
explicitly only the formula for large $t$ ($t>>\frac{1}{\nu}$)
neglecting the terms $\exp(-2\nu t)$ in comparison to the ones
without this factor
\begin{equation}
\begin{array}{l}\vert\psi_{t}\vert^{2}\simeq R^{2}L^{-1}
\exp\Big(-\nu t-\frac{15}{4}g\hbar^{2}\nu^{-3}t+
\frac{15}{4}g\nu^{-2}x^{4}\cr + \frac{45}{4}L^{-6}\hbar^{2}t^{2}\exp(-2\nu
t)\nu^{-2}x^{2}-L^{-6}\exp(-6\nu t) x^{6}\Big) .
\end{array}\end{equation}
 From eq.(137) we can see
that the sojourn time to stay in the interval $x\in [-L,L]$ which
at $g=0$ is of the order $t\simeq \nu^{-1}\ln L$ is modified by
the addition of the potential $\tilde{V}$. We know that in the
potential $V=-\frac{1}{2}\nu^{2}x^{2}+gx^{6}$ the particle located
 at $x=0$ will initially exponentially fast ($x\sim \exp(\nu t)$) depart from the maximum
at $x=0$ but this motion will be stopped by the barrier resulting
from the potential $\tilde{V}=gx^{6}$. Such a behavior is to some
extent (taking into account the roughness of our approximation)
reflected in the presence of the coupling-dependent terms in the
exponential of eq.(137) showing a slowdown at large time ($x^{4}\sim t$ ) of the departure from the
maximum of the potential. Calculation of the subsequent terms in the cumulant expansion in eq.(133)
will confirm this observation.

  \end{document}